# A BENGALI HMM BASED SPEECH SYNTHESIS SYSTEM

*Sankar Mukherjee, Shyamal Kumar Das Mandal*

Center for Educational Technology, Indian Institute of Technology Kharagpur
sankar1535@gmail.com, sdasmandal@cet.iitkgp.ernet.in

## ABSTRACT

The paper presents the capability of an HMM-based TTS system to produce Bengali speech. In this synthesis method, trajectories of speech parameters are generated from the trained Hidden Markov Models. A final speech waveform is synthesized from those speech parameters. In our experiments, spectral properties were represented by Mel Cepstrum Coefficients. Both the training and synthesis issues are investigated in this paper using annotated Bengali speech database. Experimental evaluation depicts that the developed text-to-speech system is capable of producing adequately natural speech in terms of intelligibility and intonation for Bengali.

*Index Terms*—Bengali Speech Synthesis, Text-To-Speech (TTS), Hidden-Markov-Model (HMM), Bengali HTS.

## 1. INTRODUCTION

The first requirement of a text-to-speech (TTS) system is intelligibility and the second one is the naturalness. Actually the concept of naturalness is not to restitute the reality but to suggest it. Thus, listening to a synthetic voice must allow the listener to attribute this voice to some pseudo-speaker and to perceive some kind of expressivities as well as some indices characterizing the speaking style and the particular situation of elocution [1]. Modern speech synthesizers are able to achieve high intelligibility. However, they still suffer from a rather unnatural speech. Recently, to increase the naturalness, there has been a noticeable shift from di-phone-based towards corpus-based unit selection speech synthesis observed [2]. Among these corpus-based unit selection technique is by far the best for producing the natural speech. But it requires a large database often in size of gigabyte. There are many corpus based and di-phone based TTS available for different Indian languages but those are still not reaches the acceptable quality of naturalness. More over the same has not yet been implemented for resource-limited or embedded devices such as mobile phones.

In this perspective, Hidden Markov Models (HMMs) have proven to be an efficient parametric model of the speech acoustics in the framework of speech synthesis because of its small database size and ability to produce intelligent and natural speech. Although having been originally implemented for Japanese language, the HMM-based speech synthesis (HSS) [3] approach has also been applied to other languages, e.g., English [4], German [5], Portuguese [6], Chinese [7], etc. Input contextual labels and questions for context clustering are the only language dependent topics in the HSS scheme. This paper describes first experiments on statistical parametric HMM-based speech synthesis for the Bengali language. For building of our experimental TTS system, HTS toolkit is employed.

## 2. BASIC SYSTEM

The HMM-based speech synthesis technique comprises training and synthesis parts, as depicted in Figure 1.

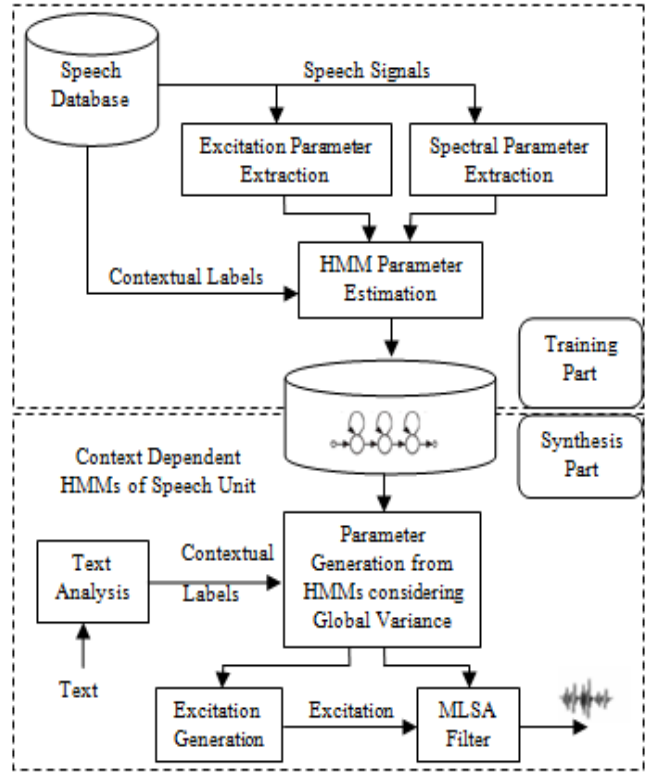

Figure-1: HMM-based speech synthesis system

## 2.1. Training

In the training part, spectrum and excitation parameters are extracted from the annotated speech database and converted to a sequence of observed feature vectors which is modeled by a corresponding sequence of HMMs. Each HMM corresponds to a left-to-right no-skip model where each output vector is composed of two streams: spectrum part, represented by mel-cepstral coefficients [8] and their related delta and delta-delta coefficients; and the excitation part, represented by Log F0 and their related delta and delta-delta coefficients. Mel-cepstral coefficients are modeled by continuous HMMs and F0s are modeled by multi-space probability distribution HMM (MSD-HMM) [9]. To capture the phonetic and prosody co-articulation phenomena Context-dependent phone models are used. State typing based on decision-tree and minimum description length (MDL) [10] criterion is applied to overcome the problem of data sparseness in training. Stream-dependent models are built to cluster the spectral, prosodic and duration features into separated decision trees.

## 2.2. Synthesis

In the synthesis phase, input text is converted first into a sequence of contextual labels through the text analysis. Then, according to such label sequence, an HMM sequence is constructed by concatenating context-dependent HMM. After this, state durations for the HMM sequence are determined so that the output probability of the state durations are maximized. Then the mel-cepstral coefficients and F0 trajectories are generated by using the parameter generation algorithm based on maximum likelihood criterion with dynamic feature and global variance constraints. Finally, speech waveform is synthesized directly from the generated mel-cepstral coefficients and F0 values by using the MLSA filter [11].

## 3. DEVELOPING BENGALI TEXT-TO-SPEECH

The speech database employed here for training purpose is originally developed by C-DAC [12]. Altogether 816 sentences are used for training which consist of 12 type of sentences i.e. Complex affirmative, Complex negative, Simple affirmative with verb, Simple affirmative without verb, Simple negative, Compound affirmative, Compound Negative, Exclamatory, Imperative, Passive, WH questions, Yes-No questions.

All the sentences were tagged with marking of phoneme, syllable, and word boundaries along with the appropriate Parts of Speech (POS) and phrase/clause markers. During the training prosodic word boundary are used as word boundary instead of syntactic word boundary. Those prosodic word labeling was carried out manually.

## 3.1. Phonemes

Bengali phoneme inventory consists of 34 consonants including two glides and 14 vowels (including 7 nasal vowels). But the occurrence of Bengali phoneme /ɲ/ is very rear hence this phoneme is not considered during the training and testing. So altogether 48 phonemes, including one silence are used for training as given in Table 1-2 along with their manner and place of articulation. All the diphthongs are marked as vowel-vowel combination.

Table-1: Bengali consonant inventory

| Phonemes | Place of articulation | Manner of articulation |
|---|---|---|
| /p/ | Bilabial | Unvoiced un-aspirated plosive/stop |
| /pʰ/ | Bilabial | Unvoiced aspirated plosive/stop |
| /b/ | Bilabial | Voiced un-aspirated plosive/stop |
| /bʰ/ | Bilabial | Voiced aspirated plosive/stop |
| /t/ | Dental | Unvoiced un-aspirated plosive/stop |
| /tʰ/ | Dental | Unvoiced aspirated plosive/stop |
| /d/ | Dental | Voiced un-aspirated plosive/stop |
| /dʰ/ | Dental | Voiced aspirated plosive/stop |
| /ʈ/ | Post-alveolar Retroflex | Unvoiced un-aspirated plosive/stop |
| /ʈʰ/ | Post-alveolar Retroflex | Unvoiced aspirated plosive/stop |
| /ɖ/ | Post-alveolar Retroflex | Voiced un-aspirated plosive/stop |
| /ɖʰ/ | Post-alveolar Retroflex | Voiced aspirated plosive/stop |
| /k/ | Velar | Unvoiced un-aspirated plosive/stop |
| /kʰ/ | Velar | Unvoiced aspirated plosive/stop |
| /g/ | Velar | Voiced un-aspirated plosive/stop |
| /gʰ/ | Velar | Voiced aspirated plosive/stop |
| /tʃ/ | Alveolar | Unvoiced un-aspirated Affricate |
| /tʃʰ/ | Alveolar | Unvoiced aspirated Affricate |
| /dʒ/ | Alveolar | Voiced un-aspirated Affricate |
| /dʒʰ/ | Alveolar | Voiced aspirated Affricate |
| /s/ | Alveolar | Unvoiced fricative |
| /ʃ/ | Post alveolar | Unvoiced fricative |
| /h/ | Glottal | Unvoiced Fricative |
| /m/ | Bilabial | Nasal murmur |
| /n/ | Dental | Nasal murmur |
| /ŋ/ | Velar | Nasal murmur |
| /ɳ/ | Palatal | Nasal retroflex murmur |
| /l/ | Dental | Lateral |
| /r/ | Alveolar | Trill |
| /ɽ/ | Post alveolar | Unaspirated Retroflex Flap |
| /ɽʰ/ | Post alveolar | Aspirated Retroflex Flap |
| /j/ | Palatal | Approximant |
| /w/ | Bilabial | Approximant |

Table-2: Bengali Vowel inventory

| Phonemes | Place of articulation | Manner of articulation |
|---|---|---|
| /u/ | Close, rounded | Back vowel |
| /o/ | Close-mid, rounded | Back vowel |
| /ɔ/ | Open, rounded | Back vowel |
| /a/ | Open, Unrounded | Central vowel |
| /æ/ | Open-mid, Unrounded | Front vowel |
| /e/ | Close-mid, Unrounded | Front vowel |
| /i/ | Close, Unrounded | Front vowel |

### 3.2. Tone

In Bengali language tone is not phonemically significant. In a simple declarative sentence with neutral focus, most words and/or phrases in Bengali is said to carry a rising tone with the exception of the last word in the sentence, which carries only a low tone. In that sense Bengali is called bound stress language. In a declarative sentence with neutral focus, the intonation pattern is falling. In sentences involving focused words or phrases, the rising tones last until the right edge of the focused word; all following words carry a low tone. WH sentences follow the same intonation pattern as the sentences involving focused words, but in Yes-No sentences the overall intonation pattern is raising. It observed that most of the bangle prosodic words whether it is spoken in sentences or isolated the F0 contours are rising [14].

To match these criteria the TOBI tone markings is done for each of the training sentences.

### 3.3. Context Based Clustering

In HMM based speech Synthesis the input contextual labels which are used to determine the corresponding HMM in the models set, depends on the language. Thus, contextual information which are fully represented in such contextual labels were necessary to be considered in order to obtain a good reproduction of the prosody.

Table 3 enumerates the main features taken into account and the main language dependent contextual factors are derived from Table 1-2. These questions represent a yes/no decision in a node of the tree. Correct questions will determine clusters to reproduce a fine F0 contour in relation to the original intonation.

### 4. SYNTHESIS OF INPUT TEXT

In order to generate the context-dependent label format from the given text first POS is marked. As existing automatic POS tagging for Bengali language is not up to the mark so it is done manually. Since the training is done based on the prosodic word, the input text prosodic word labeling is performed based on the rule as describe below [16].

Table 3: List of the context features

| Units | Features |
|---|---|
| Phoneme | -{preceding, current, succeeding} phonemes<br>-position of current phoneme in current syllable |
| Syllable | -whether or not {preceding, current, succeeding} syllables are stressed<br>-number of phonemes in {preceding, current, succeeding} syllables<br>-position of current syllable in current word<br>-number of stressed syllables in current phrase {before, after} current syllable<br>-number of syllables, counting from previous stressed to current syllable in the utterance<br>-number of syllables, counting from current to next stressed syllable in the utterance |
| Prosodic Word | -part-of-speech of {preceding, current, succeeding} words<br>-number of syllables in {preceding, current, succeeding} words<br>-position of current word in current phrase<br>-number of content words in current phrase {before, after} current word<br>-number of words counting from previous content word to current word in the utterance<br>-number of words counting from current to next content word in the utterance |
| Phrase | -number of {syllables, words} in {preceding, current, succeeding} phrases<br>-position of current phrase in current utterance<br>-TOBI endtone of current phrase |
| Utterance | -number of {syllables, words, phrases} in the utterance |

### 4.1. Prosodic Word Labeling

**Rule 1:** Hyphenated words and repeated words always form a prosodic word.
**Rule 2:** Two consecutive proper nouns, within the same prosodic phrase form a prosodic word.
**Rule 3:** If a common noun (length _ 3 syllables) is preceded by an adjective (length _ 3 syllables) then they are combined together to form a prosodic word.
**Rule 4:** A common noun and a verbal noun join together to form a prosodic word.
**Rule 5:** A postposition and the preceding word together form a prosodic word.
**Rule 6:** A verb (main or auxiliary) and the following particle together form a prosodic word.
**Rule 7:** A main verb and the following auxiliary verb (viz., a compound verb) combine together to form a prosodic word.
**Rule 8:** A common noun (or an adjective or a verbal noun) and a verb form a prosodic word.

After that the annotated text is converted to phoneme string using Grapheme to Phoneme (G2P) rules described in [13].

A complete block diagram of the above process is shown in Figure 2.

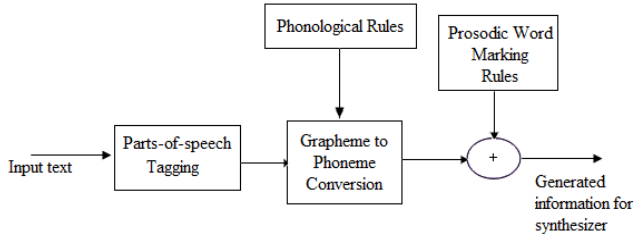

Figure 2: Block Diagram for the Linguistic Analysis of Inputted Text

## 5. EVALUATION

Figure- 3 and Figure- 4 show a comparison of spectrogram and F0 patterns between synthesized and original speech signals for a given sentence which is not included in the training database. It can be noticed that the generated spectrogram and $F_0$ contour are quite close to the natural.

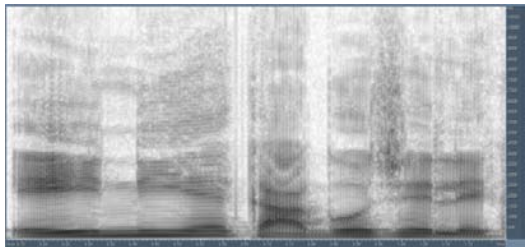
(a)

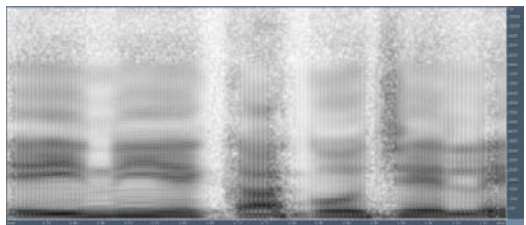
(b)

Figure- 3: Examples of spectrogram extracted from utterance "ঘনশ্যাম দাসকে বাধা পেয়ে থামতে হল" (a) Natural speech, (b) Synthesized speech

For subjective evaluation of the output speech quality 5 subjects, 3 male (L1, L2, L3) and 2 female (L4, L5), are selected and their age ranging from 24 to 50. All subjects are native speakers of Standard Colloquial Bengali and non speech expert. 10 original and synthesized sentences are randomly presented for listening and their judgment in 5 point score (1=less natural – 5=most natural).

Table 4 represents the tabulated mean opinion scores for all sentences of each subject for original as well as modified sentences.

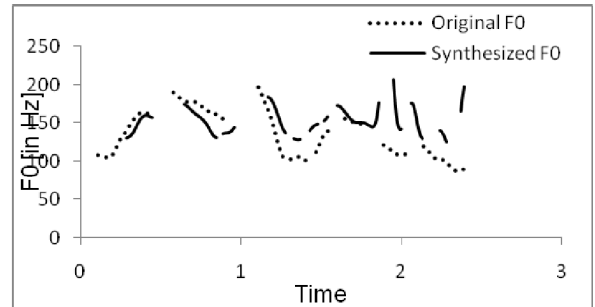

Figure- 4: Examples of $F_0$ contour extracted from utterance "ঘনশ্যাম দাসকে বাধা পেয়ে থামতে হল" (a) Natural speech, (b) Synthesized speech

Table 4 result of listing test

| Subject | | Score | | | | |
|---|---|---|---|---|---|---|
| | | L1 | L2 | L3 | L4 | L5 |
| Original Sentences | Avg | 4.9 | 4.3 | 4.7 | 4.9 | 4.5 |
| | Stdev | 0.31 | 1.05 | 0.48 | 0.31 | 0.70 |
| ESNOLA | Avg | 2.5 | 2.1 | 2.3 | 2.6 | 2.2 |
| | Stdev | 0.42 | 0.71 | 0.69 | 0.51 | 0.78 |
| HTS | Avg | 3.7 | 3.2 | 3.9 | 3.8 | 3.4 |
| | Stdev | 0.48 | 0.78 | 0.56 | 0.42 | 0.69 |

The output of the Bengali-HTS is also compare with previously developed Epoch Synchronous Non Overlap Add (ESNOLA) [15] based concatenative speech synthesis technique. The total average score for the original sentences is 4.66 and the ESNOLA based synthesis sentence is 2.34 HTS is 3.6.

## 6. CONCLUSION AND FUTURE WORKS

The evaluation results show the efficacy of HMM based Bengali TTS system for generation of highly intelligible speech with naturalness although the training corpus is not made for development of TTS system. In future the above TTS system will be trained by the appropriate training corpus for better quality of output. It was observed during the testing that the intonation of WH and Yes/no sentences was not good in spite of the presence of WH and Yes/no sentences in the training corpus. In future derived $F_0$ contour from the training model can be corrected as per the language input with the help of Fujisaki generation process $F_0$ model.